\documentclass[11pt, letterpaper, onecolumn, oneside, final]{article}

\usepackage{url}
\usepackage{times}
\usepackage{color}
\usepackage{xspace}
\usepackage{amsthm}
\usepackage{amsmath}
\usepackage{amssymb}
\usepackage{enumitem}
\usepackage[noend]{algpseudocode}
\usepackage[compact]{titlesec} 
\usepackage[font=small,labelfont=bf]{caption} 
\usepackage[left=1in,right=1in,top=1in,bottom=1in]{geometry} 
\usepackage[pdftex,bookmarks=true,bookmarksnumbered=true]{hyperref}

\newtheoremstyle{thm-compact}
{1ex} 
{1ex} 
{\itshape} 
{} 
{\bfseries} 
{.} 
{ } 
{} 
\theoremstyle{thm-compact}
\newtheorem{definition}{Definition}
\newtheorem{fact}[definition]{Fact}
\newtheorem{lemma}[definition]{Lemma}
\newtheorem{theorem}[definition]{Theorem}

\def\cgcast{\textsc{CGCast}\xspace}
\def\cseek{\textsc{CSeek}\xspace}
\def\ckseek{\textsc{CKSeek}\xspace}
\def\algcount{\textsc{Count}\xspace}

\begin{document}


\title{
	{\bf Communication Primitives in Cognitive Radio Networks}
	{\newline\Large [Regular Paper]}
}
\author{
	Seth Gilbert \\
	\small National University of Singapore \\
	\and
	Fabian Kuhn \\
	\small University of Freiburg \\
	\and
	Chaodong Zheng\footnote{Part of this research was done when Chaodong Zheng was working as a postdoctoral researcher at University of Freiburg.} \\
	\small Nanjing University \\
}
\date{
}
\maketitle
\thispagestyle{empty}

\begin{abstract}
Cognitive radio networks are a new type of multi-channel wireless network in which different nodes can have access to different sets of channels. By providing multiple channels, they improve the efficiency and reliability of wireless communication. However, the heterogeneous nature of cognitive radio networks also brings new challenges to the design and analysis of distributed algorithms.

In this paper, we focus on two fundamental problems in cognitive radio networks: neighbor discovery, and global broadcast. We consider a network containing $n$ nodes, each of which has access to $c$ channels. We assume the network has diameter $D$, and each pair of neighbors have at least $k\geq 1$, and at most $k_{max}\leq c$, shared channels. We also assume each node has at most $\Delta$ neighbors. For the neighbor discovery problem, we design a randomized algorithm $\cseek$ which has time complexity $\tilde{O}((c^2/k)+(k_{max}/k)\cdot\Delta)$. 
\cseek is flexible and robust, which allows us to use it as a generic ``filter'' to find ``well-connected'' neighbors with an even shorter running time. We then move on to the global broadcast problem, and propose \cgcast, a randomized algorithm which takes $\tilde{O}((c^2/k)+(k_{max}/k)\cdot\Delta+D\cdot\Delta)$ time. \cgcast uses \cseek to achieve communication among neighbors, and uses edge coloring to establish an efficient schedule for fast message dissemination. 

Towards the end of the paper, we give lower bounds for solving the two problems. These lower bounds demonstrate that in many situations, \cseek and \cgcast are near optimal.
\end{abstract}

\clearpage
\pagestyle{plain}
\setcounter{page}{1}


\section{Introduction}

\emph{Cognitive radio networks} are a new type of wireless network in which different devices have access to different (but overlapping) frequency bands. This asymmetry might be a result of interference (e.g., from disruptive devices or from prioritized users) or due to regulatory concerns.  
Cognitive radio networks provide a potential answer to the continuously increasing bandwidth needs of wireless communication, as they enable more graceful sharing of the available bandwidth. For example, two commonly discussed scenarios are: (1) they allow the general public to use  idle spectrum in the licensed bands that are assigned to primary users (e.g., television broadcasters)~\cite{morphy15,ieee80222}; (2) they  can increase the number of networks that can practically coexist in an unlicensed band.

Although cognitive radio networks are not a particularly new concept, our understanding remains limited when compared with the extensive research that has been done on classical single-channel or multi-channel wireless networks. In cognitive radio networks, the core challenge comes from its heterogeneous nature: different pairs of transceivers can share different sets of channels. Such unpredictable overlapping pattern (among transceivers) makes it very difficult to design and analyze efficient algorithms.  (Much interesting work in this area has focused primarily on simulations to demonstrate effectiveness, see, e.g., \cite{kondareddy08,song12,rehmani13}.) 

In this paper, we focus on two important problems: \emph{neighbor discovery}, and \emph{global broadcast}. These problems are fundamental primitives for any network.  Neighbor discovery is typically a bootstrapping procedure, and broadcast is a key building block for accomplishing many more complex tasks. We are interested in developing randomized algorithms that can solve these problems efficiently.

More specifically, we consider a cognitive radio network containing $n$ nodes, each of which has a transceiver that can access $c$ channels. If two nodes are within the transmission radius of each other and share some communication channels, then they are \emph{neighbors}. We assume all pairs of neighbors share at least $k\geq 1$ channels. We can model the network as a simple graph by letting each node denote a vertex, and connect two vertices if they are neighbors.

\paragraph{Neighbor discovery.}

To solve the neighbor discovery problem, each node must learn the identities of all of its neighbors. A simple and straightforward strategy would be for each node to randomly hop among the set of channels available to it; it would then broadcast (its identity) or listen each with some probability (e.g., using a backoff procedure to resolve contention).  This simple algorithm yields a time complexity of approximately $\tilde{O}((c^2/k)\cdot\Delta)$ in expectation, where $\Delta$ is the maximum number of neighbors a node can have.  By contrast, $\Theta(\Delta)$ is a trivial lower bound since in the worst-case, a node may receive information from only one neighbor per round (and neighbors may not be able to communicate with each other).

In this paper, we devise a new algorithm called \cseek that is much more efficient. In particular, \cseek can solve the neighbor discovery problem in $\tilde{O}((c^2/k)+(k_{max}/k)\cdot\Delta)$ time, w.h.p.\footnote{We use ``w.h.p.'' to abbreviate ``with high probability with respect to $n$'', which means an event happens with probability at least $1-1/n^\alpha$, for arbitrary constant $\alpha\geq 1$.} Here, $k_{max}$ denotes the maximum number of channels two neighbors share. Notice, if all pairs of neighbors share the same (or similar) amount of overlapping channels, then \cseek only costs $\tilde{O}(c^2/k+\Delta)$ time. As we shall later see, this nearly matches the lower bound for solving neighbor discovery in our model.

The intuition behind \cseek is that a node, instead of hopping among channels uniformly at random (e.g., in the trivial approach described above), should spend more time on channels on which it overlaps with more neighbors. Unfortunately, the pattern of overlaps is not known to the nodes in advance.  Hence, in the first part of the algorithm, nodes hop among channels uniformly at random, sampling the ``\emph{density}'' of channels; a key component of our solution in an efficient procedure for estimating density. In the second part, nodes go to dense channels more often, assuming it can hear more neighbors on these channels.  By properly dividing the time between sampling and seeking, we get an efficient algorithm.


We have also devised a variant of \cseek that can solve the \emph{$\hat{k}$-neighbor-discovery} problem. In real networks, nodes may only concern with neighbors that have strong connections with them. This motivates the $\hat{k}$-neighbor-discovery problem, in which the goal is to find (at least) all neighbors that share at least $\hat{k}\geq k$ channels with you. One variant of \cseek---which is called \ckseek---can solve the $\hat{k}$-neighbor-discovery problem in $\tilde{O}((c^2/\hat{k})+(k_{max}/\hat{k})\cdot\Delta_{\hat{k}}+\Delta)$ time. Here, $\Delta_{\hat{k}}$ is the maximum number of neighbors a node can have that share at least $\hat{k}$ channels with it. \ckseek is similar to \cseek, which demonstrates the robustness of the general strategy employed. 

\paragraph{Global broadcast.}

In this problem, a designated \emph{source} node needs to disseminate a message to all other nodes in the network. Again, one can devise a straightforward solution in which nodes hop among channels randomly and wait for the message if uninformed, or broadcast it if they are already informed. Such naive solution would cost approximately $\tilde{O}((c^2/k)\cdot D)$ time, where $D$ is the diameter of the network.

We devise a new algorithm called \cgcast that can solve global broadcast in $\tilde{O}((c^2/k)+(k_{max}/k)\cdot\Delta+D\cdot\Delta)$ time, w.h.p. Again, when $k_{max}=\Theta(k)$, the complexity of \cgcast is reduced to $\tilde{O}((c^2/k)+D\cdot\Delta)$. 

The performance improvements of \cgcast comes from two design decisions. First, for an informed node to quickly disseminate the message to its neighbors, we do an edge coloring and use the solution to establish an efficient deterministic schedule. Secondly, in order to solve edge coloring efficiently, we use \cseek as a primitive for communication. This again demonstrates the flexibility and robustness of \cseek.

\paragraph{Lower bounds.}

In the last part of the paper, we devise lower bounds for solving neighbor discovery and global broadcast in the model we considered. (The techniques here originally derive from~\cite{Newport14}.)

More specifically, to prove the $\Omega(c^2/k+\Delta)$ lower bound for neighbor discovery, we consider a combinatorial game which captures the core difficulty of the problem, and then do a reduction argument to prove our claim. The combinatorial game has been previously used to prove other lower bounds in cognitive radio networks (see, e.g., \cite{gilbert15}), but the reduction argument is tailored for our purpose.

On the other hand, we also show a $\Omega((c^2/k)+D\cdot\min\{c,\Delta\})$ lower bound for solving global broadcast. This proof is quite straightforward, and we include it mainly for the sake of completeness.

As can be seen, for \cseek, when $k_{max}=\Theta(k)$, it matches the lower bound within poly-logarithmic factor; and for \cgcast, when $k_{max}=\Theta(k)$ and $c\geq\Delta$, again our bounds are near tight.

\section{Related Work}

There is much ongoing work in the area of cognitive radio networks, both in terms of low-level implementation issues (e.g.,~\cite{akyildiz08} and~\cite{yucek09}) and in terms of algorithms.  In this paper, we focus on the question of how to design algorithms for these types of networks.


\paragraph{Neighbor discovery.}

In traditional single-channel or multi-channel wireless networks, neighbor discovery is a problem that has been extensively studied under a variety of different models (w.r.t., e.g., synchrony and availability of collision detection). Interested readers can, e.g., refer to the survey paper from Khan et al.~\cite{khan15}, and from Chen et al.~\cite{chen16}, for more details. On the contrary, for cognitive radio networks, limited focus has been put on this problem. (See related parts in \cite{khan15} for a survey.)  We are only able to identify one work from Zeng et al.~\cite{zeng16}\footnote{Despite the journal version~\cite{zeng16} was published in 2016, the original conference version~\cite{mittal11} appeared in 2011.} that explicitly targets this issue and provides theoretical results. 

In their interesting paper, the authors design and analyze several algorithms for neighbor discovery in cognitive radio networks under a variety of assumptions for both synchronous and asynchronous systems. In particular, using our terminology/model, their algorithm takes $\tilde{O}(c^2/k+c\cdot\Delta/k)$ time. Since $c \geq k_{max}$, our solution is always at least as fast as theirs (asymptotically).


We also note here that neighbor discovery can potentially be solved by using other algorithms that were originally designed for other purposes. For example, solutions to rendezvous~\cite{shin10,lin11,gu13,chen14,gu14}---a popular problem which requires each pair of neighbors in the network to meet every so often---can be helpful. The only problem is, contention may exist when meeting happens, thus simple meeting does not alway imply successful exchange of identities. Indeed, most recent (also shown to be near optimal) results~\cite{chen14,gu14} can solve rendezvous in $\tilde{O}(c^2)$ time. When ignoring contention and assuming $k=O(1)$, this bound matches the simple algorithm we described in the introduction section.\footnote{The algorithms described in \cite{chen14,gu14} are deterministic, while the simple one we described earlier is randomized.} The difficult part, and what \cseek achieves, is to resolve contention when meeting happens, and maintain low time complexity in the meantime.

\paragraph{Broadcast.}

The goal of a broadcast is to distribute information in a network. Typically, there are two kinds of broadcast: local broadcast, in which the goal is to disseminate the message to your immediate neighbors; and global broadcast, in which all nodes in the network need to receive the message.

In the cognitive radio network setting, there are a few papers addressing the broadcast problem; the models and approaches employed are highly diverse. For example, in a series of nice papers by Song and Xie~\cite{song12,song14,song15}, by carefully constructing channel hopping sequences, the authors are able to accomplish broadcast in multi-hop cognitive radio networks. These works analyze the probability that a broadcast will be successful, but do not provide a theoretical guarantee on broadcast time, instead focusing on simulations to show the effectiveness. In this paper, we take a full-on algorithmic approach, guaranteeing successful global broadcast with high probability and analyzing our algorithms for worst-case performance. For the local broadcast problem, in a paper by Gilbert et al.~\cite{gilbert15}, the authors propose a simple randomized algorithm that is shown to be near optimal. In another paper by Kondareddy et al.~\cite{kondareddy08}, the authors use ``minimal neighbor graph'' to achieve local broadcast, which in turn helps to disseminate control information. Finally, another approach to tackle the broadcast problem is to assume that all necessary information (e.g., channel availability, network topology) is known in advance, and then focus on developing efficient algorithms which can find good broadcast schedules. Both \cite{arachchige11} and \cite{ji13} belong to this category. (In this paper, we assume minimal a priori environmental knowledge.)

\section{Model}


We consider a synchronous wireless network containing $n$ nodes, each of which has a unique identity. 

Each node has a radio transceiver that can access $c$ channels, and different nodes can potentially access different sets of channels. Moreover, we assume for each channel, different nodes may have different labels for it. That is, we do not assume a \emph{global channel label} exist. (Instead, nodes have a \emph{local channel label}.)

For two nodes, if they are within the transmission range of each other and share some channels, then they are \emph{neighbors}. For each pair of neighboring nodes, we assume they share at least $k\geq 1$ channels, and at most $k_{max}\leq c$ channels. For each node, initially, it does not know the identities of its neighbors, it also does not know the set of channels on which it can communicate with its neighbors.

Now, we can model the network as a simple graph $\mathcal{G}$, in which each vertex denotes a node, and there is an edge connecting two nodes iff they are neighbors. We assume $\mathcal{G}$ is connected, and has diameter $D$. We also assume the maximum degree of $\mathcal{G}$ is $\Delta$, which implies a node can have at most $\Delta$ neighbors.

We divide time into discrete \emph{slots}. In each time slot, each radio transceiver can only operate on one of the $c$ available channels. For a node $u$, if it decides to broadcast a message in a time slot, then it only ``receives'' that message in that slot. If $u$ decides to listen on a channel and no neighbors of $u$ broadcast on that channel in that slot, then $u$ hears nothing (i.e., silence). If $u$ decides to listen on a channel and among the neighbors of $u$, exactly one broadcasts on that channel in that slot, then $u$ hears the message from that node. Finally, if $u$ decides to listen on a channel and multiple neighbors of $u$ broadcast on that channel in that slot, then $u$ hears nothing. That is, nodes cannot distinguish between silence and multiple nodes broadcasting simultaneously. (Alternatively, we can say we assume collision detection is \emph{not} available.)

Finally, we assume nodes start execution simultaneously, and can independently generate random bits.

\section{Algorithms for Neighbor Discovery}

In this section, we introduce \cseek, a randomized algorithm that can be used to solve neighbor discovery efficiently. We will also present a variant of \cseek that can be used to solve $\hat{k}$-neighbor-discovery.

\subsection{\algcount: A Counting Procedure}

Before presenting \cseek and \ckseek, we first introduce a simple counting algorithm \algcount that allows a node to get an estimate on how many other nodes are on the same channel. In particular, the estimate given by \algcount is guaranteed to be within a small (multiplicative) constant factor of the actual value. Algorithm \algcount is generic and can easily be plugged into other cognitive radio network algorithms.

More specifically, \algcount solves the following problem: On a channel, there is one listening node and an unknown number of broadcasters. The listener wants to learn the count of broadcasters. Notice, since the degree upper bound $\Delta$ is known, nodes know that the actual number of broadcasters cannot exceed $\Delta$.

To solve this problem, \algcount needs $\lg{\Delta}$ rounds, each of which contains $\Theta(\lg{n})$ time slots. Hence, the total time complexity is $O(\lg^2{n})$. At a high level, the algorithm takes the classical ``guess and verify'' approach. More specifically, in each round, nodes will have an estimate about the actual count. The estimate starts from one, and doubles after each round. In each round, each broadcaster will broadcast with probability that is proportional to the reciprocal of the estimate, and the listener will count how many messages it can receive within this round. When the estimate is close to the actual count, the number of messages received by the listener should reach a certain peak value. Hence, the listener can obtain a relatively accurate count.

Due to space constraint, detailed description and analysis of \algcount is left in Appendix \ref{sec-app-count}. Here, we only states the guarantees provided by it.

\begin{lemma}\label{lemma-count}
The counting procedure \algcount takes $O(\lg^2{n})$ time slots, and allows the listener to obtain a count that is in $[m,4m]$, w.h.p. Here, $m$ is the actual number of broadcasters.
\end{lemma}

\subsection{The \cseek Algorithm}

We now present the \cseek algorithm, which contains two parts. Intuitively, it is designed in the following way. In part one, each node samples all the channels that are available to it to get an idea about how many of its neighbors have access to each channel. Moreover, as we shall later see, for a node $u$, part one also offers an opportunity for it to know the identities of the neighbors that overlap with it on less crowded channels (i.e., channels on which $u$ does not overlap with many neighbors). In part two, $u$ will go to more crowded channels more often, so that neighbors that overlap with it on these channels---even though there may exist many such neighbors---get enough chances to introduce themselves to $u$.

More specifically, part one contains $\Theta((c^2/k)\cdot\lg{n})$ steps, each of which contains $O(\lg^2{n})$ slots. Hence, the total time complexity of part one is $O((c^2/k)\cdot\lg^3{n})$. In each step, each node will go to one of the $c$ channels that is available to it uniformly at random, and then choose to be a broadcaster or listener each with probability $1/2$. Then, within current step, each node will run the counting procedure \algcount on the selected channel. Listeners will record the count, and any identities they have heard. Broadcasters, on the other hand, will broadcast their identities when necessary (i.e., according to \algcount.)

The second part of \cseek contains $\Theta((k_{max}/k)\cdot\Delta\cdot\lg{n})$ steps, each of which contains $\lg{\Delta}$ slots. Hence, the total time complexity of part two is $O((k_{max}/k)\cdot\Delta\cdot\lg^2{n})$. In each step, each node will choose to be a broadcaster or listener each with probability $1/2$. If a node $u$ chooses to be a broadcaster, then it will pick one of the $c$ available channels uniformly at random and go to that channel. However, if $u$ chooses to be a listener, then it will pick an available channel according to the total count it has obtained during part one. More specifically, if the total count $u$ has obtained for channel $ch$ during part one is $x_{ch}$, then it will choose $ch$ in a step in part two with probability $x_{ch}/\sum_{ch'\in\mathcal{C}_{u}}{x_{ch'}}$. Here, $\mathcal{C}_{u}$ is the set of channels that are available to $u$. Once nodes have picked channels, in a step, a listener will listen in every slot and record any identity it has heard; a broadcaster, on the other hand, will broadcast its identity with probability $2^{i-1}/\Delta$ in the $i$\textsuperscript{th} slot of current step. The pseudocode of \cseek is shown in Figure \ref{fig-cseek} in Appendix \ref{sec-app-code}.

\subsection{Analysis of the \cseek Algorithm}


Consider a node $u$ and one of its neighbors $v$. The first key technical result states that if for most of the channels on which $u$ and $v$ overlap, at most $O(c)$ neighbors of $u$ overlap with $u$ on each of these channels, then by the end of part one, $u$ will know $v$'s identity. That is, if most of the channels $u$ and $v$ share are not too crowded, then $u$ will know $v$'s identity after part one. More precisely, we have the following lemma.

\begin{lemma}\label{lemma-cseek-part-one}
Assume $v$ is one of $u$'s neighbors, also assume $u$ and $v$ overlap on $k_{u,v}$ channels. Assume among these channels, there exist at least $k_{u,v}/2$ channels such that for each of these channels, at most $8c$ of $u$'s neighbors overlap with $u$ on this channel. Then, during part one, after $O((c^2/k_{u,v})\cdot\lg{n})$ steps each of which containing $O(\lg^2{n})$ slots, $u$ will hear $v$'s identity, w.h.p.
\end{lemma}

We now sketch the proof of Lemma \ref{lemma-cseek-part-one}.\footnote{Due to space constraint, if not otherwise stated, detailed proofs of lemmas and theorems are provided in Appendix \ref{sec-app-proof}.} In each step in part one, for $u$ to hear $v$'s identity, three conditions must hold: (a) $u$ is a listener and $v$ is a broadcaster; (b) $u$ and $v$ choose the same channel; and (c) $v$ broadcasts alone in some slot during the counting procedure. According to protocol description, obviously condition (a) holds with probability $1/4$. Moreover, it is not hard to see that condition (b) holds with probability $k_{u,v}/c^2$. Hence, the tricky part is to calculate the probability that condition (c) holds. Assume (a) and (b) indeed happen, we know there is at least a fifty percent chance that for the channel $u$ and $v$ have chosen, there are at most $8c$ neighbors of $u$ that can access this channel. If this is indeed the case, then in expectation, at most $(8c-1)\cdot(1/c)\cdot(1/2)=O(1)$ neighbors of $u$ (beside $v$) are playing as broadcasters on this channel in this step. That is, there is only a \emph{constant} number of broadcasters. Since the counting procedure \algcount that is executed within the step contains a rounds in which the estimate (approximately) matches the number of broadcasters, and since each round in \algcount contains $\Theta(\lg{n})$ slots, this ``constant level contention'' can be fully resolved. Therefore, conditioned on (a) and (b) indeed happen, (c) will happen with at least some constant probability. As a result, we can conclude in each step, $u$ will hear $v$'s identity with probability at least $\Omega(k_{u,v}/c^2)$. Since each step is independent, the lemma follows.

In the second key technical result, we consider the case in which most of the channels on which $u$ and $v$ overlap are crowded. (I.e., the ``complement'' of the previous case.) In this scenario, we claim that $u$ will hear $v$'s identity during part two.

\begin{lemma}\label{lemma-cseek-part-two}
Assume $v$ is one of $u$'s neighbors, also assume $u$ and $v$ overlap on $k_{u,v}$ channels. Assume among these channels, there exist at least $k_{u,v}/2$ channels such that for each of these channels, at least $8c$ of $u$'s neighbors overlap with $u$ on this channel. Then, during part two, after $O((\sum_{w\in\mathcal{N}_u}{k_{u,w}})/k_{u,v}\cdot\lg{n})$ steps each of which containing $\lg{\Delta}$ slots, $u$ will hear $v$'s identity, w.h.p. Here, $\mathcal{N}_u$ denotes the set of $u$'s neighbors.
\end{lemma}

The proof for Lemma \ref{lemma-cseek-part-two} is more involved than that of Lemma \ref{lemma-cseek-part-one}. Again, we only provide the sketch here.

To begin with, notice that during part one, for node $u$ and each of its neighbor $v$, they have met (in the sense that they choose the same channel while $u$ is listener and $v$ is broadcaster) in $\Theta((k_{u,v}/k)\cdot\lg{n})$ steps. This implies the sum of the counts (for all channels $u$ can access) $u$ \emph{should} obtain is $\Theta(\sum_{w\in\mathcal{N}_{u}}{(k_{u,w}/k)}\cdot\lg{n})$. Since the counting procedure can provide estimate within constant factor of error, we know the actual sum $u$ obtained is also $\Theta(\sum_{w\in\mathcal{N}_{u}}{(k_{u,w}/k)}\cdot\lg{n})$.

Now, we calculate the probability that $u$ and $v$ choose the same \emph{crowded} channel $ch$ in one step during part two when $u$ is listener and $v$ is broadcaster. Here, ``crowded'' means at least $8c$ of $u$'s neighbors overlap with $u$ on this channel. According to the protocol, we know the probability is $(x_{ch}/(\Theta(\sum_{w\in\mathcal{N}_{u}}{(k_{u,w}/k)}\cdot\lg{n})))\cdot(1/c)=\Theta(x_{ch}/(c\cdot\lg{n}\cdot\sum_{w\in\mathcal{N}_{u}}{(k_{u,w}/k)}))$. Here, $x_{ch}$ is the count obtained by $u$ for channel $ch$. The next step is to calculate $x_{ch}$, and it is not hard to prove the expectation of $x_{ch}$ is $\Theta(n_{ch}/k\cdot\lg{n})$. Here, $n_{ch}$ is the number of neighbors that overlap with $u$ on channel $ch$. Since channel $ch$ is crowded, we know $n_{ch}\geq 8c\geq 8k$. By using a standard Chernoff bound~\cite{mitzenmacher05}, we can prove \emph{with high probability} $x_{ch}$ is $\Theta(n_{ch}/k\cdot\lg{n})$. At this point, we can claim in each step during part two, with probability at least $\Theta(x_{ch}/(c\cdot\lg{n}\cdot\sum_{w\in\mathcal{N}_{u}}{(k_{u,w}/k)}))=\Theta(n_{ch}/(c\cdot\sum_{w\in\mathcal{N}_{u}}{k_{u,w}}))$: node $u$ and $v$ will both choose a crowded channel $ch$, while $u$ is a listener and $v$ is a broadcaster.

To let $u$ hear $v$'s identity, we still need to take contention on channel $ch$ into consideration. It is not hard to show, in expectation, there are $(n_{ch}-1)\cdot(1/2)\cdot(1/c)=\Theta(n_{ch}/c)$ neighbors (beside $v$) that are playing as broadcaster on $ch$. Again, since $n_{ch}\geq 8c$, by using a Chernoff bound, we can prove that with at least some constant probability, there are $\Theta(n_{ch}/c)$ other neighbors of $u$ that will broadcast on $ch$.

Since there is a back-off procedure within each step in part two, we can now conclude: in one step in part two, with probability at least $\Theta(n_{ch}/(c\cdot\sum_{w\in\mathcal{N}_{u}}{k_{u,w}}))\cdot\Theta(c/n_{ch})=\Theta(1/\sum_{w\in\mathcal{N}_{u}}{k_{u,w}})$, $u$ will hear $v$'s identity on a specific crowded channel $ch$. Since there are at least $k_{u,v}/2$ such crowded channels, in each step in part two, the probability that $u$ will hear $v$'s identity is at least $(k_{u,v}/2)\cdot\Omega(1/\sum_{w\in\mathcal{N}_{u}}{k_{u,w}})=\Omega(k_{u,v}/\sum_{w\in\mathcal{N}_{u}}{k_{u,w}})$. Since each step in part two is independent, the lemma follows.

Combine Lemma \ref{lemma-cseek-part-one} and Lemma \ref{lemma-cseek-part-two} will immediately lead to the following theorem.

\begin{theorem}\label{thm-cseek}
\cseek can solve the neighbor discovery problem in $O((c^2/k)\cdot\lg^3{n}+(k_{max}/k)\cdot\Delta\cdot\lg^2{n})=\tilde{O}((c^2/k)+(k_{max}/k)\cdot\Delta)$ time slots, w.h.p.
\end{theorem}

\subsection{\ckseek: Using \cseek as a General Filter}

The \cseek algorithm allows nodes to discover all neighbors. However, sometimes, it may be desirable to discover only ``well connected'' neighbors. In particular, it is possible that a node may only want to find neighbors that overlap with it on sufficiently many channels. In this part, we show that the \cseek algorithm can solve this problem as well, even with a shorter running time! 

To be concrete, we consider the $\hat{k}$-neighbor-discovery problem, in which the goal is to let each node find (at least) all neighbors that overlap with it on at least $\hat{k}\geq k$ channels. Moreover, for the ease of presentation, for a node $u$, if a neighbor $v$ overlaps with it on at least $\hat{k}$ channels, then $v$ is a \emph{good neighbor} of $u$.

To solve the $\hat{k}$-neighbor-discovery problem, we only need to make small adjustments to the original \cseek algorithm. In particular, in the modified version---which is called \ckseek---part one contains only $\Theta((c^2/\hat{k})\cdot\lg{n})$ steps, and part two contains only $\Theta(((k_{max}/\hat{k})\cdot\Delta_{\hat{k}}+\Delta+c)\cdot\lg{n})$ steps. Here, $\Delta_{\hat{k}}$ denotes the maximum number of good neighbors a node can have. Notice, if an estimate of $\Delta_{\hat{k}}$ is not available, we can simply run part two longer, making it containing $\Theta(((k_{max}/\hat{k})\cdot\Delta+c)\cdot\lg{n})$ steps.

Consider a node $u$ and one of its good neighbor $v$, we now prove the correctness of \ckseek.

Similar to the analysis of \cseek, we still consider two complement cases. In the first case, we claim that if for at least half of the channels on which $u$ and $v$ overlap, at most $8c$ neighbors of $u$ overlap with it on each of these channels, then by the end of part one of \ckseek, $u$ will know $v$'s identity, w.h.p. This is a direct application of Lemma \ref{lemma-cseek-part-one}.

In the second case, we assume most of the channels $u$ and $v$ share are crowded. In such scenario, $u$ will know $v$'s identity during part two of \ckseek. More specifically, we have the following lemma.

\begin{lemma}\label{lemma-ckseek-part-two}
Assume $v$ is one of $u$'s good neighbors, also assume $u$ and $v$ overlap on $k_{u,v}$ channels. Assume among these channels, there exist at least $k_{u,v}/2$ channels such that for each of these channels, at least $8c$ of $u$'s neighbors overlap with $u$ on this channel. Then, during part two of \ckseek, after $O(((k_{max}/\hat{k})\cdot\Delta_{\hat{k}}+\Delta+c)\cdot\lg{n})$ steps each of which containing $\lg{\Delta}$ slots, $u$ will hear $v$'s identity, w.h.p.
\end{lemma}

Lemma \ref{lemma-ckseek-part-two} looks very similar to Lemma \ref{lemma-cseek-part-two}, and it is tempting to use the same idea to prove it. At a high level, this is indeed the case: we still calculate the probability that $u$ hears $v$'s identity in one step during part two. However, the detailed procedure and technique is quite different. Among the differences, the most significant one is how to bound the sum of the counts $u$ obtained during part one. In $\cseek$, part one is sufficiently long, and we can claim during part one $u$ will meet each of its neighbor $v$ for $\Theta((k_{u,v}/k)\cdot\lg{n})$ times, in expectation, and w.h.p. This immediately tells us the counts sum to $\Theta(\sum_{w\in\mathcal{N}_{u}}{(k_{u,w}/k)}\cdot\lg{n})$, w.h.p. In \ckseek, however, part one is shorter, and we can only claim during part one $u$ will meet its neighbor $v$ for $\Theta((k_{u,v}/\hat{k})\cdot\lg{n})$ times in expectation, but not w.h.p. To overcome this difficulty, we come up with a more careful and complex way to calculate the sum, allowing us to bound it by $O(((k_{max}/\hat{k})\cdot\Delta_{\hat{k}}+\Delta+c)\cdot\lg{n})$, w.h.p. Again, due to space constraint, we leave the full proof to the appendix.

Based on our above analysis, we can easily obtain the following theorem. Notice, it implies when $\hat{k}>k$, the (asymptotic) runtime of \ckseek is strictly shorter than that of \cseek.

\begin{theorem}\label{thm-nd-filter}
\ckseek can solve the $\hat{k}$-neighbor-discovery problem in $O((c^2/\hat{k})\cdot\lg^3{n}+((k_{max}/\hat{k})\cdot\Delta_{\hat{k}}+\Delta)\cdot\lg^2{n})=\tilde{O}((c^2/\hat{k})+(k_{max}/\hat{k})\cdot\Delta_{\hat{k}}+\Delta)$ time slots, w.h.p. 
\end{theorem}

\section{Algorithm for Global Broadcast}

In this part, we introduce \cgcast, an efficient algorithm to solve global broadcast in cognitive radio networks. We will begin with an overview, then proceed to the details, and finally show its correctness.

\subsection{Overview}

The key reason that the simple global broadcast algorithm described in the introduction section is inefficient is that nodes are \emph{randomly} hopping among channels, making it slow to propagate the message among neighbors. To solve this problem, in \cgcast, we let neighbors establish a \emph{deterministic} schedule of meeting, so that once a node is informed, it can quickly disseminate the message.

More specifically, to establish a communication schedule among neighbors, we reduce the problem to \emph{edge coloring}~\cite{peleg00}. In particular, for each node, it will color each edge that is connected to it from a set of $\Theta(\Delta)$ colors, and at the same time guaranteeing each edge has a unique color. Once this is done, we can imagine a broadcast protocol which proceeds in steps, each of which contains $\Theta(\Delta)$ rounds. By mapping each round in a step to a color, each pair of neighbors have a ``dedicated'' round for communication in each step. In this way, informed nodes can quickly disseminate the message to their neighbors.

However, to solve edge coloring, more efforts are needed. First, by using \emph{line graph}~\cite{peleg00}, we can reduce edge coloring to node coloring (i.e., vertex coloring), in which a node and all of its neighbors are quired to have unique colors. There are many efficient algorithms for this task. However, one problem is that these algorithms usually assume a node can send a potentially different message to every neighbor in one time slot. This is not true in our model. To solve this issue, we need to leverage the neighbor discovery algorithm developed in previous sections. In particular, notice that to solve neighbor discovery, a node needs to receive some information from each of its neighbors. Therefore, if we can solve neighbor discovery in $T$ time, then we can use the same algorithm to allow each pair of neighbors to exchange one message in $T$ time.

\subsection{The \cgcast Algorithm}

We now proceed to describe the details of the \cgcast algorithm, which can be divided into three main parts: the neighbor discovery part, the edge coloring part, and the message dissemination part.

Firstly, we will first run the \cseek neighbor discovery algorithm, so that each node can know its neighbors. This process will take $\tilde{O}((c^2/k)+(k_{max}/k)\cdot\Delta)$ time slots, as Theorem \ref{thm-cseek} suggests.

Once neighbor discovery is done, we will work on the line graph of the original network graph $\mathcal{G}$, so that we can solve edge coloring in $\mathcal{G}$ by solving node coloring in its line graph. Notice, the line graph $\mathcal{G}_L$ of $\mathcal{G}$ is defined in the following way: each edge $(u,v)$ in $\mathcal{G}$ is a node $w_{u,v}$ in $\mathcal{G}_L$, and two nodes in $\mathcal{G}_L$ are connected if and only if they share one common endpoint in $\mathcal{G}$. Also notice, in the line graph $\mathcal{G}_L$, for a (virtual) node $(u,v)$, it is simulated by the (physical) node in $\mathcal{G}$ which has smaller identity. Since each node in $\mathcal{G}$ knows the identities of its neighbors, such simulation can be correctly done in a consistent manner.

We now describe how to obtain a valid node coloring of $\mathcal{G}_L$, which is an implementation of the algorithm described in \cite{luby93} in our model.\footnote{There exist better node coloring algorithms which are faster, or use less colors. However, these algorithms are more complex, and usually do not provide significantly better results (for our setting). Moreover, node coloring is not the focus of this paper. Therefore, we only use a simple and relatively efficient algorithm here.} The node coloring procedure contains $\Theta(\lg{n})$ phases, each of which contains two steps, and each step contains $\tilde{\Theta}((c^2/k)+(k_{max}/k)\cdot\Delta)$ time slots. Initially, each (virtual) node in $\mathcal{G}_L$ has an identical color plate containing $2\Delta$ different colors, and all nodes in $\mathcal{G}_L$ are active. At the beginning of the first step of each phase, for each active node in $\mathcal{G}_L$, with probability $1/2$, it chooses to do nothing in this step. Otherwise, it randomly chooses a color that is still available in its color plate with uniform probability. Then, during the $\tilde{\Theta}((c^2/k)+(k_{max}/k)\cdot\Delta)$ time slots within current step, neighboring active nodes in $\mathcal{G}_L$ will exchange the colors they have chosen. (This is possible as one execution of \cseek allows neighboring nodes in $\mathcal{G}$ to exchange information once. Moreover, for any two physical nodes simulating two neighboring virtual nodes in $\mathcal{G}_L$, they are at most two hops away.) If two active neighboring nodes in $\mathcal{G}_L$ have chosen the same color, then both of them will give up their choices. Otherwise, they will keep their choices. In the second step within current phase, neighboring active nodes in $\mathcal{G}_L$ will use the $\tilde{\Theta}((c^2/k)+(k_{max}/k)\cdot\Delta)$ time slots to exchange their choices on the colors. If an active node has decided on a color, then it will become inactive after current phase. Otherwise, an active node will remember which colors have already been chosen by its neighbors, remove these colors, and then continue into the next phase.

Once the node coloring procedure is done, we need to run \cseek one more time, so that for each virtual node $w_{u,v}$ in $\mathcal{G}_L$, the physical node (in $\mathcal{G}$) which simulates this virtual node can inform the choice of color (of edge $(u,v)$) to the other physical node. At this time point, we have a $2\Delta$ edge coloring of $\mathcal{G}$.

The next, and final part would be to leverage the edge coloring to quickly disseminate the message. The message dissemination part contains $D$ phases, each of which contains $2\Delta$ steps. Each step has $\Theta(\lg{n})$ rounds, and each round costs $\lg{\Delta}$ time slots. Therefore, the total time complexity of the message dissemination part is $O(D\cdot\Delta\cdot\lg^2{n})=\tilde{O}(D\cdot\Delta)$ time slots. The message dissemination part proceeds in the following way. Nodes map the $2\Delta$ colors to the $2\Delta$ steps within each phase according to some predefined rule. Moreover, for each pair of neighboring nodes, among the set of channels that is available to both of them, they can fix a special ``dedicated communication channel'' during the initial neighbor discovery part. (We discuss how to fix such special channels in the next paragraph.) Now, consider a node $u$. In each phase, in the $i$\textsuperscript{th} step, assume the corresponding color is $\mathcal{K}$. If $u$ is not adjacent to an edge with color $\mathcal{K}$, then it will stay idle in this step. Otherwise, assume edge $(u,v)$ has color $\mathcal{K}$. In such case, in this step, $u$ will go to the special channel that it has agreed with $v$. Now, if $u$ does not know the message yet, it will listen during this step. Otherwise, if $u$ is already informed, then within each of the $\Theta(\lg{n})$ rounds in current step, it will do a back-off style broadcast on the chosen channel. (This is why each round costs $\lg{\Delta}$ time slots). 

Before proceeding to the analysis, we describe one possible method which allows two neighbors to fix a dedicated communication channel despite the absence of global channel label. During the initial neighbor discovery part, when $u$ hears $v$ for the first time, it records the number of the current slot. Then, after the neighbor discovery part, and before the edge coloring part, nodes will run \cseek once more. This time, in the message, $u$ includes not only its identity, but also the numbers of time slots in which it heard its neighbors for the first time in the previous execution of \cseek. Now, for a pair of neighboring nodes $u$ and $v$, assume $u$ hears $v$ for the first time in slot $t_{u,v}$, and $v$ hears $u$ for the first time in slot $t_{v,u}$. Then, $u$ and $v$ will fix the channel they used in slot $\min\{t_{u,v},t_{v,u}\}$ as their dedicated communication channel. Since all nodes start executing \cgcast simultaneously, we know both $u$ and $v$ can correctly determine $\min\{t_{u,v},t_{v,u}\}$.

\subsection{Analysis of the \cgcast Algorithm}

To show the correctness of \cgcast, firstly notice that a valid $2\Delta$ node coloring of $\mathcal{G}_L$ gives a valid $2\Delta$ edge coloring for network graph $\mathcal{G}$. This follows from the definition of line graph.

\begin{fact}\label{fact-gb-line-graph-convert}
For a simple graph $\mathcal{G}$ with maximum degree $\Delta$, if we have a valid $2\Delta$ node coloring for its line graph $\mathcal{G}_L$, then we can obtain a valid $2\Delta$ edge coloring for $\mathcal{G}$.
\end{fact}

We then prove our algorithm can correctly generate a $2\Delta$ node coloring for line graph $\mathcal{G}_L$. In particular, we adopt the proof in Section 3.3 of \cite{luby93} to our model. The high level idea of the proof is that after each phase, with some constant probability, a constant fraction of the remaining nodes in $\mathcal{G}_L$ will decide their colors and become inactive. As a result, within $O(\lg{n})$ phases, all nodes in $\mathcal{G}_L$ will terminate, w.h.p.

\begin{lemma}\label{lemma-gb-node-color}
In \cgcast, after executing the coloring procedure, we have a $2\Delta$ node coloring of $\mathcal{G}_L$, w.h.p.
\end{lemma}

The last step is to show the effectiveness of the message dissemination part. In essence, after each phase, the message will propagate one hop. Since the network has diameter $D$, we know all nodes in the network will know the message after $D$ phases. This concludes our analysis and leads to the following theorem.

\begin{theorem}\label{thm-cgcast}
\cgcast can solve the global broadcast problem in $O((c^2/k)\cdot\lg^4{n}+(k_{max}/k)\cdot\Delta\cdot\lg^3{n}+D\cdot\Delta\cdot\lg^2{n})=\tilde{O}((c^2/k)+(k_{max}/k)\cdot\Delta+D\cdot\Delta)$ time slots, w.h.p.
\end{theorem}

\section{Lower Bounds}

\subsection{Lower Bound for Neighbor Discovery}

In this part, we will derive a $\Omega((c^2/k)+\Delta)$ lower bound for the neighbor discovery problem in cognitive radio networks. It demonstrates \cseek is near optimal (within poly-logarithmic factor) when the number of overlapping channels for different pairs of neighbors are same or similar (i.e., $k_{max}=\Theta(k)$).

To begin with, we argue the $\Omega(\Delta)$ factor in the lower bound. Imagine a star network and consider the node at the center. Call this node $u$. In each time slot, we know $u$ can receive at most one message. Moreover, in each message, there is at most one neighbor's identity. This is due to the topology of the network. Hence, we know at least $\Delta$ time slots is needed for $u$ to learn all neighbors' identities.

We then focus on the $\Omega(c^2/k)$ factor, which needs more efforts. Our strategy is to first derive a lower bound for winning a combinatorial game, and then connect the game to the neighbor discovery problem by a reduction argument. We have previously studied this game and used it to derive lower bounds for the local broadcast problem in cognitive radio networks~\cite{gilbert15}. Here, we will reuse some of the results.

To obtain the needed $\Omega(c^2/k)$ bound, the game we considered is slightly different when $k\leq c/2$ and $c/2<k\leq c$. We first focus on the $k\leq c/2$ scenario.

\paragraph{The $k\leq c/2$ scenario.} In this setting, we consider the \emph{$(c,k)$-bipartite hitting game}. In this game, the input is two integers $c, k$ such that $1\leq k\leq c/\beta$, where $\beta\geq 2$ is a constant. Consider two sets of nodes each of size $c$: $A=\{a_1,a_2,\cdots,a_c\}$ and $B=\{b_1,b_2,\cdots,b_c\}$. Let $\mathcal{H}$ be a complete bipartite graph on bipartition $(A,B)$. The game is played between a \emph{player} and a \emph{referee}. At the beginning of the game the referee privately selects a matching $M$ of size $k$ from $\mathcal{H}$. The game then proceeds in rounds. In each round, the player \emph{proposes} an edge $e$ in $\mathcal{H}$. If $e\in M$ the player \emph{wins}, otherwise it moves on to the next round.

In the lemma below, a lower bound is shown for winning the $(c,k)$-bipartite hitting game.

\begin{lemma}[Lemma 11 from \cite{gilbert15}]\label{lemma-nd-lower-bound-game1}
Let $\mathcal{P}$ be a player that guarantees to win the $(c,k)$-bipartite hitting game in $f(c,k)$ rounds with probability at least $1/2$, for some $1\leq k\leq c/\beta$, and some constant $\beta\geq 2$. It follows that $f(c,k)\geq c^2/(\alpha k)=\Theta(c^2/k)$, where $2<\alpha=2(\beta/(\beta-1))^2\leq 8$.
\end{lemma}

We then reduce $(c,k)$-bipartite hitting to neighbor discovery. In particular, we construct a network containing only two nodes and then simulate an algorithm $\mathcal{A}$ running in it. We demonstrate that if $\mathcal{A}$ can achieve neighbor discovery fast, then we can use the simulation process to solve $(c,k)$-bipartite hitting fast. We note that the reduction strategy we used here is different from the one that appeared in \cite{gilbert15}.

\begin{lemma}\label{lemma-nd-lower-bound-reduction}
Let $\mathcal{A}$ be an algorithm that guarantees to solve neighbor discovery in $g(c,k,n)$ slots with probability at least $1/2$. Then, one can use $\mathcal{A}$ to construct a player $\mathcal{P}_{\mathcal{A}}$ that guarantees to win the $(c,k)$-bipartite hitting game in $O(g(c,k,n))$ rounds, for any $1\leq k\leq c$, with probability at least $1/2$.
\end{lemma}

\paragraph{The $k>c/2$ scenario.} In this setting, we use a similar strategy as the $k\leq c/2$ scenario, but focusing on a different hitting game. In particular, we consider the \emph{$c$-complete bipartite hitting game}. The input to this game is an integer $c\geq 1$. Consider two sets of nodes $A=\{a_1,a_2,\cdots,a_c\}$ and $B=\{b_1,b_2,\cdots,b_c\}$. Let $\mathcal{H}$ be a complete bipartite graph on bipartition $(A,B)$. The game is again played between a \emph{player} and a \emph{referee}. At the beginning of the game the referee privately selects a \emph{maximum} matching $M$ in $\mathcal{H}$. The game then proceeds in rounds. In each round, the player \emph{proposes} an edge $e$ in $\mathcal{H}$. If $e\in M$ the player \emph{wins}, otherwise it moves on to the next round. Again, we can show a lower bound for winning this hitting game.

\begin{lemma}[Lemma 14 from \cite{gilbert15}]\label{lemma-nd-lower-bound-game2}
Let $\mathcal{P}$ be a player that guarantees to win the $c$-complete bipartite hitting game in $f(c)$ rounds with probability at least $1/2$, for some positive integer $c$. It follows that $f(c)\geq c/3$.
\end{lemma}

Now, notice Lemma \ref{lemma-nd-lower-bound-reduction} is also applicable to the $k>c/2$ scenario (as the $(c,k)$-bipartite hitting game becomes the $c$-complete bipartite hitting game when $k=c$). This implies there exists a strategy for using a fast neighbor discovery algorithm to solve $c$-complete bipartite hitting fast. Therefore, by combining Lemma \ref{lemma-nd-lower-bound-game1}, \ref{lemma-nd-lower-bound-game2}, and Lemma \ref{lemma-nd-lower-bound-reduction}, along with the $\Omega(\Delta)$ lower bound we previously discussed, Theorem \ref{thm-nd-lower-bound} follows.

\begin{theorem}\label{thm-nd-lower-bound}
For any algorithm, to solve the neighbor discovery problem under the considered model with probability at least $1/2$, the time consumption is at least $\Omega((c^2/k)+\Delta)$.
\end{theorem}

\subsection{Lower Bounds for Global Broadcast}

In this part, we give a $\Omega((c^2/k)+D\cdot\min\{c,\Delta\})$ lower bound for solving global broadcast in cognitive radio networks. More precisely, we have the following theorem.

\begin{theorem}\label{thm-gb-lower-bound}
For any algorithm, to solve the global broadcast problem under the considered model with probability at least $1/2$, the time consumption is at least $\Omega((c^2/k)+D\cdot\min\{c,\Delta\})$.
\end{theorem}

Proving this theorem is not too difficult. The argument for the $\Omega(c^2/k)$ part is similar to the one we used when proving the lower bound for neighbor discovery. In particular, we can reduce the $(c,k)$-bipartite hitting game discussed earlier to two node broadcast (i.e., one source node and one uninformed node). For the $\Omega(D\cdot\min\{c,\Delta\})$ part, we consider the case where the network graph $\mathcal{G}$ is a complete tree with each non-leaf node having $\min\{c,\Delta\}-1$ children. Then, for the message to propagate one hop, we show it will take at least $\Theta(\min\{c,\Delta\})$ time slots. Since the network diameter is $D$, we can obtain the bound as desired.

\section{Summary and Discussion}

In this paper, we focus on providing algorithms that can act as key communication primitives in cognitive radio networks. In particular, we propose several randomized algorithms that can solve the neighbor discovery problem and the global broadcast problem efficiently. Unlike many previous work, our algorithms provide provable performance guarantees, and we show these upper bounds are near optimal in many cases by deriving corresponding lower bounds. Nevertheless, we suspect there is still room for improvements, from either the algorithmic side or the lower bound side.

Consider the neighbor discovery problem as an example, the upper and lower bound can still have a big gap when different pairs of neighbors share significantly different number of channels. This is a result of how we design \cseek. In particular, in part two of \cseek, for a node $u$, the algorithm gives priority to the more crowded channels. On the other hand, for nodes that overlap with $u$ on more channels, during part one, $u$ will meet them more often, creating the impression that certain channels are more crowded. Moreover, in the back-off procedure within each step in part two, all nodes are competing fairly. As a result of these design decisions, during part two, node $u$ is more likely to hear nodes that overlap with it on many channels. This slows down the algorithm's progress. To solve this issue, one possible way is to somehow give priority to nodes that overlap with $u$ on less channels. However, this is not easy, as node $u$ has no information about these neighbors before discovering them. More importantly, even if $u$ knows a neighbor $v$ overlaps with it on many channels, it may not be feasible to simply ask $v$ to reduce it's broadcasting probability, as another neighbor of $v$ may only overlap with $v$ on few channels. Alternatively, we may try to improve the lower bound based on these observations. However, this again seems highly non-trivial.


\clearpage
\bibliographystyle{abbrv}
\bibliography{draft-podc-1}

\clearpage
\appendix
\section*{Appendix}

\section{Detailed Description and Analysis of \algcount}\label{sec-app-count}

We now describe the counting algorithm in more detail. As mentioned previously, the counting procedure contains $\lg{\Delta}$ rounds, each of which contains $\Theta(\lg{n})$ time slots. In round $i$, the nodes will estimate the count of broadcasters is $2^{i-1}$, and validate if this estimate is accurate or not. More specifically, in each round, the listener will simply listen in each slot, and count the number of messages that it has heard during this round. On the other hand, for each broadcaster, in round $i$, in each time slot, it will broadcast its identity with probability $1/2^{i-1}$, and do nothing otherwise. After a round $i$, if, for the first time since the start of the first round, the listener has heard messages in more than $(1+\delta)\cdot(8e^{-7})$ fraction of slots in this round, then the listener will use $2^{i+1}$ as the count. Here, $0<\delta<1$ is a sufficiently small constant.

We now show \algcount can provide a relatively accurate count.

Assume the actual number of broadcasters is $m\leq\Delta$. In round $i$, in a slot, the probability that the listener will hear a message is $m\cdot(1/2^{i-1})\cdot(1-1/2^{i-1})^{m-1}$. Hence, if the length of a round is $a\lg{n}$ where $a$ is a sufficiently large constant, then in expectation, the listener will hear $a\cdot m\cdot(1/2^{i-1})\cdot(1-1/2^{i-1})^{m-1}\cdot\lg{n}$ messages in round $i$. Let random variable $X_i$ denote this number, thus $\mathbb{E}(X_i)=a\cdot m\cdot(1/2^{i-1})\cdot(1-1/2^{i-1})^{m-1}\cdot\lg{n}$.

Define function $f(x)=a\cdot m\cdot x\cdot(1-x)^{m-1}\cdot\lg{n}$. (Here, $x$ is effectively representing $1/2^{i-1}$.) We know the first order derivative of $f(x)$ is $f'(x)=(a\cdot m\cdot(1-x)^{m-2}\cdot\lg{n})\cdot(1-x\cdot m)$. This implies, when $xm=1$, function $f(x)$ will have a maximum value. In turn, this suggests, when $2^{i-1}=m$, the value of $\mathbb{E}(X_i)$ will be maximum.

Therefore, when $1\leq 2^{i-1}\leq m/8$, we know $\mathbb{E}(X_i)\leq a\cdot m\cdot(8/m)\cdot(1-8/m)^{m-1}\cdot\lg{n}=8a\lg{n}\cdot(1-8/m)^{m-1}\leq 8a\lg{n}\cdot e^{-8(m-1)/m}\leq (a\lg{n})\cdot(8e^{-8(7m/8)/m})\leq(a\lg{n})\cdot(8e^{-7})$. Since each slot is independent, apply a Chernoff bound and we know, when $2^{i-1}\leq m/8$, in one round, the maximum fraction of rounds in which the listener can hear a message is $(1+\delta)\cdot(8e^{-7})$, w.h.p. Here, when $a$ is sufficiently large, $0<\delta<1$ is an arbitrarily small constant.

On the other hand, when $m\geq 2^{i-1}\geq m/2\geq 2$, we know $\mathbb{E}(X_i)\geq a\cdot m\cdot(2/m)\cdot(1-2/m)^{m-1}\cdot\lg{n}=2a\lg{n}\cdot(1-2/m)^{m-1}\geq 2a\lg{n}\cdot e^{-4(m-1)/m}\geq (a\lg{n})\cdot(2e^{-4})$. Since each slot is independent, apply a Chernoff bound and we know, when $m\geq 2^{i-1}\geq m/2$, in one round, the minimum fraction of rounds in which the listener can hear messages is $(1-\delta)\cdot(2e^{-4})$, w.h.p.

Notice, our counting algorithm asks the listener to use the estimate in the first round in which the fraction of clear message slots is more than $(1+\delta)\cdot(8e^{-7})$. Therefore, we now know the listener will not use the count when $2^{i-1}\leq m/8$, and must have already obtained the count when $2^{i-1}\geq m$. (This is because $8e^{-7}<2e^{-4}$.) Therefore, we know the listener will obtain a count that is within $[m,4m]$, w.h.p.

\section{Pseudocode of Algorithms and Procedures}\label{sec-app-code}

See Figure \ref{fig-cseek} on page \pageref{fig-cseek}.

\begin{figure}[!ht]
\hrule
\vspace{1ex}\textbf{Pseudocode of \cseek executed at node $u$:}\vspace{1ex}
\hrule
\begin{small}
\begin{algorithmic}[1]
\State $sum\gets 0, ids\gets\emptyset$
\For {($i=1$ to $c$)} $counts[i]\gets 0$ \Comment $counts$ is a dictionary used to store the count for each channel \EndFor
\Statex \textsc{Part I}
\For {($i=1$ to $\Theta((c^2/k)\cdot\lg{n})$)}
	\State $ch\gets\texttt{random}(1,c)$ \Comment Choose channel, where $\texttt{random}(x,y)$ returns a random integer in $[x,y]$
	\State $role\gets\texttt{random}(0,1)$ \Comment Choose to be broadcaster or listener
	\State $\langle count_{ch}, ids_{ch}\rangle\gets\algcount(ch,role)$ \Comment Execute \algcount with specified role on specific channel
	\State $sum\gets sum+count_{ch}, counts[ch]\gets counts[ch]+count_{ch}$ \Comment Update counts
	\State $ids\gets ids\cup ids_{ch}$ \Comment Update identities
\EndFor
\Statex \textsc{Part II}
\For {($i=1$ to $\Theta((k_{max}/k)\cdot\Delta\cdot\lg{n})$)}
	\State $role\gets\texttt{random}(0,1)$ \Comment Choose to be broadcaster or listener
	\If {($role==0$)} \Comment If $u$ is a broadcaster
		\State $ch\gets\texttt{random}(1,c)$ \Comment Choose channel
		\For {($j=\lg{\Delta}$ to $1$)} \Comment Do back-off style broadcast
			\If {($\texttt{random}(1,2^{j})==1$)} $\texttt{broadcast}(ch,id_{u})$ \Comment $id_{u}$ is the identity of $u$ \EndIf
		\EndFor
	\Else \Comment If $u$ is a listener
		\State $rnd\gets\texttt{random}(1,sum), ch=1$
		\While {($rnd>counts[ch]$)} \Comment Choose channel $ch$ with probability $count_{ch}/\sum_{ch'\in\mathcal{C}_{u}}{count_{ch'}}$
			\State $rnd\gets rnd-counts[ch], ch\gets ch+1$
		\EndWhile
		\For {($j=\lg{\Delta}$ to $1$)} \Comment Listen on the chosen channel and record identities
			\State $id\gets\texttt{listen}(ch), ids\gets ids\cup\{id\}$
		\EndFor
	\EndIf
\EndFor
\Statex \textsc{Return Result}
\State \textbf{return} $ids$
\end{algorithmic}
\end{small}
\hrule\vspace{1ex}
\caption{Pseudocode of the \cseek algorithm.}\label{fig-cseek}
\vspace{-3ex}
\end{figure}

\section{Proofs for Lemmas and Theorems}\label{sec-app-proof}

\begin{proof}[\bf{Proof of Lemma \ref{lemma-cseek-part-one}}]
Assume the set of channels on which $u$ and $v$ overlap is $\mathcal{K}_{u,v}$. By assumption, we know $|\mathcal{K}_{u,v}|=k_{u,v}$. Let $\mathcal{K}'_{u,v}$ be a subset of $\mathcal{K}_{u,v}$ which is of size at least $k_{u,v}/2$, such that for each channel in $\mathcal{K}'_{u,v}$, node $u$ overlaps with at most $8c$ of its neighbors. 

We calculate the probability that $u$ will hear $v$'s identity in a step during part one. In particular, we consider the case in which $u$ hear $v$'s identity on one of the channels in $\mathcal{K}'_{u,v}$ in one step. For this to happen, $u$ must choose to be a listener and $v$ must choose to be a broadcaster, which happens with probability $(1/2)\cdot(1/2)=1/4$. Moreover, $u$ must choose a channel in $\mathcal{K}'_{u,v}$, and $v$ must choose that same channel as well, this happens with probability at least $(k_{u,v}/2c)\cdot(1/c)=k_{u,v}/(2c^2)$.

The next key factor that needs to be taken into consideration is contention: other neighbors of $u$ may choose to broadcast on this channel as well. To quantify the level of contention, consider another neighbor of $u$ that overlaps with $u$ on the channel on which $u$ and $v$ have chosen. Call this neighbor $w$, and the channel chosen by $u$ and $v$ as $ch$. For $w$ to broadcast on channel $ch$: (a) $w$ must choose to be a broadcaster, which happens with probability $1/2$; and (b) $w$ must choose channel $ch$, which happens with probability $1/c$. Notice, we assume at most $8c$ neighbors of $u$ overlap with $u$ on channel $ch$. Hence, we know beside $v$, in expectation, at most $(8c-1)\cdot(1/2)\cdot(1/c)\leq 4$ other nodes will broadcast on channel $ch$. Apply Markov's inequality, we know with probability at least $1/2$, there are at most eight other neighbors of $u$ will broadcast on channel $ch$ in this step.

As this point, we can conclude: during part one, in a step, with probability at least $(1/4)\cdot(k_{u,v}/(2c^2))\cdot(1/2)=\Theta(k_{u,v}/c^2)$: (a) $u$ and $v$ will choose the same channel $ch$; (b) $u$ will listen and $v$ will broadcast; and (c) there are at most eight other broadcasters on channel $ch$.

Now, assume the above good event indeed happens, notice the counting procedure in each step. In round four, the estimate will be eight, and each broadcaster executing the counting procedure will broadcast with probability $1/8$. Hence, in each slot in round four, there is a constant probability that $v$ will be the sole broadcaster, and its identity will be received by $u$. Therefore, by the end of this round, $v$'s identity will be heard by $u$, w.h.p.

As a result, we can now conclude, in one step, $u$ will hear $v$'s identity with probability at least $\Theta(k_{u,v}/c^2)$. Since each step is independent, we can immediately have the lemma.
\end{proof}

\begin{proof}[\bf{Proof of Lemma \ref{lemma-cseek-part-two}}]
Assume node $u$ has $\Delta_{u}$ neighbors, call this set of neighbors $\mathcal{N}_{u}$. Assume the set of channels on which $u$ and $v$ overlap is $\mathcal{K}_{u,v}$. Let $\mathcal{K}'_{u,v}$ be a subset of $\mathcal{K}_{u,v}$ which is of size at least $k_{u,v}/2$, such that for each channel in $\mathcal{K}'_{u,v}$, node $u$ overlaps with at least $8c$ of its neighbors.

Before proving the lemma, we show some simple facts.

First of all, notice that after part one of protocol execution, for a channel $ch$, if the count obtained by $u$ is $x_{ch}$, then the actual count $\hat{x}_{ch}$ is in range of $[x_{ch}/4,4x_{ch}]$, w.h.p. This is because, according to Lemma \ref{lemma-count}, the counting procedure can provide a relatively accurate count that is within constant factor of error.

The above fact further implies another conclusion: w.h.p.\ $\sum_{ch\in\mathcal{C}_{u}}{x_{ch}}=\Theta(\sum_{w\in\mathcal{N}_{u}}{(k_{u,w}/k)}\cdot\lg{n})$, where $\mathcal{C}_{u}$ is the set of channels that is available to $u$. This is because, in one step in part one, for $u$ to meet a neighbor $w$ (but not necessarily hear $w$'s identity) while $u$ is a listener and $w$ is a broadcaster, the probability is $(1/2)\cdot(1/2)\cdot(k_{u,w}/c)\cdot(1/c)=k_{u,w}/(4c^2)$. Since part one contains $\Theta((c^2/k)\cdot\lg{n})$ steps, we know $u$ will meet $w$ while $u$ is a listener and $w$ is a broadcaster for $\Theta((k_{u,w}/k)\cdot\lg{n})$ times, w.h.p. Therefore, we know $\sum_{ch\in\mathcal{C}_{u}}{\hat{x}_{ch}}=\Theta(\sum_{w\in\mathcal{N}_{u}}{(k_{u,w}/k)}\cdot\lg{n})$, w.h.p. Since $\sum_{ch\in\mathcal{C}_{u}}{x_{ch}}$ is a constant factor estimate of $\sum_{ch\in\mathcal{C}_{u}}{\hat{x}_{ch}}$, we can conclude $\sum_{ch\in\mathcal{C}_{u}}{x_{ch}}=\Theta(\sum_{w\in\mathcal{N}_{u}}{(k_{u,w}/k)}\cdot\lg{n})$, w.h.p.

With the above facts, we now proceed to prove the lemma.

Consider a step in part two of protocol execution, we calculate the probability that $u$ hear $v$'s identity on one of the channels in $\mathcal{K}'_{u,v}$ in this step.

To calculate this probability, we first consider the probability that $u$ meets $v$ on a specific channel $ch\in\mathcal{K}'_{u,v}$ while $u$ is a listener and $v$ is a broadcaster. For this to happen, first, $u$ must choose to listen and $v$ must choose to broadcast, which happens with probability $1/4$. Then, $u$ and $v$ must choose the same channel $ch$ that is in $\mathcal{K}'_{u,v}$, this happens with probability $(x_{ch}/(\Theta(\sum_{w\in\mathcal{N}_{u}}{(k_{u,w}/k)}\cdot\lg{n})))\cdot(1/c)=\Theta(x_{ch}/(c\cdot\lg{n}\cdot\sum_{w\in\mathcal{N}_{u}}{(k_{u,w}/k)}))$.

We now calculate the value of $x_{ch}$. In one step in part one, for $u$ and one of its neighbor $w$ which overlaps with $u$ on channel $ch$, the probability that $u$ meets $w$ on channel $ch$ while $u$ is a listener and $w$ is a broadcaster is $(1/2)\cdot(1/2)\cdot(1/c)\cdot(1/c)=1/4c^2$. Hence, in expectation, the actual count $u$ should have is $n_{ch}\cdot(1/4c^2)\cdot\Theta((c^2/k)\cdot\lg{n})=\Theta(n_{ch}/k\cdot\lg{n})$. Here, $n_{ch}$ is the number of neighbors that overlap with $u$ on channel $ch$. Notice, since $ch$ is a channel in $\mathcal{K}'_{u,v}$, we know $n_{ch}\geq 8c\geq 8k$. Moreover, in part one, nodes make choices independently in each step, and each node makes choices independently in different steps. Hence, apply a Chernoff bound, we know the actual count $u$ should have is $\Theta(n_{ch}/k\cdot\lg{n})$, w.h.p. Since the counting procedure can provide constant factor estimate, we know $x_{ch}=\Theta(n_{ch}/k\cdot\lg{n})$, w.h.p.

At this point, we know in one step in part two, with probability at least $\Theta(x_{ch}/(c\cdot\lg{n}\cdot\sum_{w\in\mathcal{N}_{u}}{(k_{u,w}/k)}))=\Theta(n_{ch}/(c\cdot\sum_{w\in\mathcal{N}_{u}}{k_{u,w}}))$: node $u$ and $v$ will both choose channel $ch\in\mathcal{K}'_{u,v}$, and $u$ will be a listener, and $v$ will be a broadcaster.

The next key factor that needs to be taken into consideration is the level of contention, as other broadcaster may try to broadcast on channel $ch$ as well. To quantify such contention, consider another neighbor of $u$ which also overlaps with $u$ on channel $ch$. We call this neighbor $w$, and calculate the probability that $w$ broadcasts on $ch$ in this step. For this to happen: (a) $w$ must choose to broadcast, which happens with probability $1/2$; and (b) $w$ must choose channel $ch$, which happens with probability $1/c$. Notice, we have assumed there are $n_{ch}\geq 8c$ neighbors of $u$ that overlap with it on channel $ch$. Hence, in expectation, beside $v$, in one step in part two, there will be $(n_{ch}-1)\cdot(1/2)\cdot(1/c)$ other broadcasters on channel $ch$. Notice, $3\leq 3n_{ch}/(8c)\leq(n_{ch}-1)\cdot(1/2)\cdot(1/c)\leq n_{ch}/(2c)$, and each node makes choices independently. Hence, apply a Chernoff bound and we know, with at least some constant probability, there are $\Theta(n_{ch}/c)$ other neighbors of $u$ that will broadcast on channel $ch$ in one step during part two.

Since there is a back-off procedure within each step in part two, we can now conclude: in one step in part two, with probability at least $\Theta(n_{ch}/(c\cdot\sum_{w\in\mathcal{N}_{u}}{k_{u,w}}))\cdot\Theta(c/n_{ch})=\Theta(1/\sum_{w\in\mathcal{N}_{u}}{k_{u,w}})$, $u$ will hear $v$'s identity on channel $ch\in\mathcal{K}'_{u,v}$. Define $e_{ch_{i}}$ to be the event that $u$ hears $v$'s identity on channel $ch_{i}\in\mathcal{K}'_{u,v}$, we know $\mathbb{P}(e_{ch_i})=\Omega(1/\sum_{w\in\mathcal{N}_{u}}{k_{u,w}})$. Notice, $\{e_{ch_i}|ch_i\in\mathcal{K}'_{u,v}\}$ is a set of disjoint events. Moreover, we have assumed $|\mathcal{K}'_{u,v}|\geq k_{u,v}/2$. Therefore, in one step in part two, the probability that $u$ will hear $v$'s identity is at least $(k_{u,v}/2)\cdot\Omega(1/\sum_{w\in\mathcal{N}_{u}}{k_{u,w}})=\Omega(k_{u,v}/\sum_{w\in\mathcal{N}_{u}}{k_{u,w}})$. Since each step in part two is independent, we know after $\Theta((\sum_{w\in\mathcal{N}_u}{k_{u,w}})/k_{u,v}\cdot\lg{n})$ steps, $u$ will know $v$'s identity w.h.p.
\end{proof}

\begin{proof}[\bf{Proof of Lemma \ref{lemma-ckseek-part-two}}]
Assume node $u$ has $\Delta_{u}$ neighbors, call this set of neighbors $\mathcal{N}_{u}$. Assume $u$ has $\hat{\Delta}_{u}$ good neighbors, call this set of neighbors $\hat{\mathcal{N}}_{u}$. Assume the set of channels $u$ and $v$ share is $\mathcal{K}_{u,v}$. Let $\mathcal{K}'_{u,v}$ be a subset of $\mathcal{K}_{u,v}$ which is of size at least $k_{u,v}/2$, such that for each channel in $\mathcal{K}'_{u,v}$, node $u$ overlaps with at least $8c$ of its neighbors. Assume $\mathcal{C}_{u}$ is the set of channels that is available to $u$, for a channel $ch\in\mathcal{C}_u$, let $n_{ch}$ denote the number of neighbors that share this channel with $u$. Let $\mathcal{C}'_u$ be a subset of $\mathcal{C}_u$ such that for each $ch\in\mathcal{C}'_u$ we have $n_{ch}\geq\hat{k}$.

Before proving the lemma, we show an important claim: w.h.p., we have $\sum_{ch\in\mathcal{C}_{u}}{x_{ch}}=O(((k_{max}/\hat{k})\cdot\Delta_{\hat{k}}+\Delta+c)\cdot\lg{n})$. Here, $x_{ch}$ is the sum of the counts node $u$ obtained for channel $ch$ during part one. To prove this, consider a specific channel $ch$, and calculate the value of $x_{ch}$. During part one, in each step, node $u$ will choose a channel uniformly at random and go to that channel. Since part one contains $\Theta((c^2/\hat{k})\cdot\lg{n})$ steps, by using a Chernoff bound, we know during part one, $u$ will listen on channel $ch$ for $\Theta((c/\hat{k})\cdot\lg{n})$ steps, w.h.p. For now, let us focus on these $\Theta((c/\hat{k})\cdot\lg{n})$ steps. Define indicator random variable $y_{i,w}$ to denote whether node $w$---which is a neighbor of $u$---has chosen to broadcast on channel $ch$ during the $i$\textsuperscript{th} step of these $\Theta((c/\hat{k})\cdot\lg{n})$ steps. We know $x_{ch}=\Theta(\sum_{i}\sum_{w\in\mathcal{N}_{u}}{y_{i,w}})$ because $\sum_{i}\sum_{w\in\mathcal{N}_{u}}{y_{i,w}}$ is the \emph{actual} count for $ch$, and Lemma \ref{lemma-count} shows the counting procedure within each step can provide a constant factor estimate of the actual count.

We now try to calculate $\sum_{i}\sum_{w\in\mathcal{N}_{u}}{y_{i,w}}$. Consider a neighbor $w$ of $u$ that share channel $ch$ with $u$, consider a step $i$ among the $\Theta((c/\hat{k})\cdot\lg{n})$ steps in which $u$ chooses to listen on channel $ch$. According to our protocol, for $y_{i,w}$ to be one, node $w$ must choose the same channel with $u$ and must choose to broadcast. Hence, we know $\mathbb{P}(y_{i,w}=1)=(1/2)\cdot(1/c)=\Theta(1/c)$. Therefore, we know $\mathbb{E}(\sum_{i}\sum_{w\in\mathcal{N}_{u}}{y_{i,w}})=n_{ch}\cdot\Theta((c/\hat{k})\cdot\lg{n})\cdot\Theta(1/c)=\Theta((n_{ch}/\hat{k})\cdot\lg{n})$.\footnote{Notice, we use $n_{ch}$, not $\Delta_u$, in the expression, as it can be the case that not all neighbors of $u$ share channel $ch$ with $u$.} Notice, it is easy to see that for different values of $i$ and/or $w$, random variables $y_{i,w}$ are independent. Hence, apply a standard Chernoff bound, we know when $n_{ch}\geq\hat{k}$, we have $\sum_{i}\sum_{w\in\mathcal{N}_{u}}{y_{i,w}}=\Theta((n_{ch}/\hat{k})\cdot\lg{n})$, w.h.p. Similarly, when $n_{ch}<\hat{k}$, apply a variant of Chernoff bound (see Theorem 4.4 of \cite{mitzenmacher05}), we have $\sum_{i}\sum_{w\in\mathcal{N}_{u}}{y_{i,w}}=O(\lg{n})$, w.h.p.

Recall $\mathcal{C}'_u$ denotes a subset of $\mathcal{C}_u$ such that for each $ch\in\mathcal{C}'_u$ we have $n_{ch}\geq\hat{k}$. At this point, we know w.h.p.\ $\sum_{ch\in\mathcal{C}_{u}}{x_{ch}}=\sum_{ch\in\mathcal{C}'_u}{\Theta((n_{ch}/\hat{k})\cdot\lg{n})}+\sum_{ch\in\mathcal{C}_u\backslash\mathcal{C}'_u}{O(\lg{n})}\leq\sum_{ch\in\mathcal{C}_u}{\Theta((n_{ch}/\hat{k})\cdot\lg{n})}+\sum_{ch\in\mathcal{C}_u\backslash\mathcal{C}'_u}{O(\lg{n})}\leq\sum_{ch\in\mathcal{C}_u}{\Theta((n_{ch}/\hat{k})\cdot\lg{n})}+c\cdot O(\lg{n})$. Notice that $\sum_{ch\in\mathcal{C}_u}{n_{ch}}=\sum_{w\in\mathcal{N}_u}{k_{u,w}}\leq\Delta_{\hat{k}}\cdot k_{max}+(\Delta-\Delta_{\hat{k}})\cdot\hat{k}\leq\Delta_{\hat{k}}\cdot k_{max}+\Delta\cdot\hat{k}$. Thus, we know w.h.p.\ $\sum_{ch\in\mathcal{C}_{u}}{x_{ch}}\leq\sum_{ch\in\mathcal{C}_u}{\Theta((n_{ch}/\hat{k})\cdot\lg{n})}+c\cdot O(\lg{n})\leq\Theta(((k_{max}/\hat{k})\cdot\Delta_{\hat{k}}+\Delta)\cdot\lg{n})+O(c\cdot\lg{n})=O(((k_{max}/\hat{k})\cdot\Delta_{\hat{k}}+\Delta+c)\cdot\lg{n})$. This proves our claim.

We are now ready to prove the lemma. In particular, consider a step in part two, we calculate the probability that $u$ hears $v$'s identity on one of the channels in $\mathcal{K}'_{u,v}$ in this step.

To calculate this probability, we first consider the probability that $u$ meets $v$ on a specific channel $ch\in\mathcal{K}'_{u,v}$ while $u$ is a listener and $v$ is a broadcaster. For this to happen, first, $u$ must choose to listen and $v$ must choose to broadcast, which happens with probability $1/4$. Then, $u$ and $v$ must both choose $ch\in\mathcal{K}'_{u,v}$, this happens with probability $(x_{ch}/O(((k_{max}/\hat{k})\cdot\Delta_{\hat{k}}+\Delta+c)\cdot\lg{n}))\cdot(1/c)=\Omega(x_{ch}/(((k_{max}/\hat{k})\cdot\Delta_{\hat{k}}+\Delta+c)\cdot c\cdot\lg{n}))$.

Since the channel $ch$ we are considering is in $\mathcal{K}'_{u,v}$, we know $n_{ch}\geq 8c>\hat{k}$. Hence, according to our previous analysis, we know $x_{ch}=\Theta((n_{ch}/\hat{k})\cdot\lg{n})$, w.h.p. As a result, we know in one step in part two, with probability at least $\Omega(x_{ch}/(((k_{max}/\hat{k})\cdot\Delta_{\hat{k}}+\Delta+c)\cdot c\cdot\lg{n}))=\Omega(n_{ch}/((k_{max}\cdot\Delta_{\hat{k}}+\hat{k}\cdot\Delta+\hat{k}\cdot c)\cdot c))$: node $u$ and $v$ will both choose channel $ch\in\mathcal{K}'_{u,v}$, and $u$ will be a listener, and $v$ will be a broadcaster.

The next key factor that needs to be taken into consideration is the level of contention, as other broadcaster may try to broadcast on channel $ch$ as well. To quantify such contention, consider another neighbor of $u$ which also overlaps with $u$ on channel $ch$. We call this neighbor $w$, and calculate the probability that $w$ broadcasts on $ch$ in this step. For this to happen: (a) $w$ must choose to broadcast, which happens with probability $1/2$; and (b) $w$ must choose channel $ch$, which happens with probability $1/c$. Notice, we have assumed there are $n_{ch}\geq 8c$ neighbors of $u$ that overlap with it on channel $ch$. Hence, in expectation, beside $v$, in one step in part two, there will be $(n_{ch}-1)\cdot(1/2)\cdot(1/c)$ other broadcasters on channel $ch$. Notice, $3\leq 3n_{ch}/(8c)\leq(n_{ch}-1)\cdot(1/2)\cdot(1/c)\leq n_{ch}/(2c)$, and each node make choices independently. Hence, apply a Chernoff bound and we know, with at least some constant probability, there are $\Theta(n_{ch}/c)$ other neighbors of $u$ that will broadcast on channel $ch$ in one step during part two.

Since there is a back-off procedure within each step in part two, we can now conclude: in one step in part two, with probability at least $\Omega(n_{ch}/((k_{max}\cdot\Delta_{\hat{k}}+\hat{k}\cdot\Delta+\hat{k}\cdot c)\cdot c))\cdot\Theta(c/n_{ch})=\Omega(1/(k_{max}\cdot\Delta_{\hat{k}}+\hat{k}\cdot\Delta+\hat{k}\cdot c))$, $u$ will hear $v$'s identity on channel $ch\in\mathcal{K}'_{u,v}$. Define $e_{ch_{i}}$ to be the event that $u$ hears $v$'s identity on channel $ch_{i}\in\mathcal{K}'_{u,v}$, we know $\mathbb{P}(e_{ch_i})=\Omega(1/(k_{max}\cdot\Delta_{\hat{k}}+\hat{k}\cdot\Delta+\hat{k}\cdot c))$. Notice, $\{e_{ch_i}|ch_i\in\mathcal{K}'_{u,v}\}$ is a set of disjoint events. Moreover, we have assumed $|\mathcal{K}'_{u,v}|\geq k_{u,v}/2$. Therefore, in one step in part two, the probability that $u$ will hear $v$'s identity is at least $(k_{u,v}/2)\cdot\Omega(1/(k_{max}\cdot\Delta_{\hat{k}}+\hat{k}\cdot\Delta+\hat{k}\cdot c))=\Omega(\hat{k}/(k_{max}\cdot\Delta_{\hat{k}}+\hat{k}\cdot\Delta+\hat{k}\cdot c))=\Omega(1/((k_{max}/\hat{k})\cdot\Delta_{\hat{k}}+\Delta+c))$. Since each step in part two is independent, we know within $O(((k_{max}/\hat{k})\cdot\Delta_{\hat{k}}+\Delta+c)\cdot\lg{n})$ steps, $u$ will know $v$'s identity w.h.p.
\end{proof}

\begin{proof}[\bf{Proof of Fact \ref{fact-gb-line-graph-convert}}]
According to the definition of line graph, each node in $\mathcal{G}_L$ denotes an edge in $\mathcal{G}$, hence by giving each node in $\mathcal{G}_L$ a color, we have effectively given each edge in $\mathcal{G}$ a color. Moreover, for any two edges that share same endpoints in $\mathcal{G}$, they must each denote a node in $\mathcal{G}_L$, and these two nodes must be connected. If the node coloring in $\mathcal{G}_L$ is valid, then these two edges must have different colors in $\mathcal{G}$, which implies the edge coloring in $\mathcal{G}$ is valid as well.
\end{proof}

\begin{proof}[\bf{Proof of Lemma \ref{lemma-gb-node-color}}]
There are several points worth noting before we prove the lemma.

Firstly, notice that in one execution of \cseek, each pair of neighboring nodes can exchange one piece of information. This is a direct corollary of Theorem \ref{thm-cseek}.

Secondly, consider two neighboring (virtual) nodes in $\mathcal{G}_L$: node $w_{u_1,u_2}$ and node $w_{u_1,u_3}$. Here, $w_{u_1,u_2}$ denotes the edge $(u_1,u_2)$ in $\mathcal{G}$ and $w_{u_1,u_3}$ denotes the edge $(u_1,u_3)$ in $\mathcal{G}$. If these two virtual nodes are simulated by the same physical node (in particular, $u_1$), then for every step in each phase of the coloring procedure, they can (locally) exchange one piece of information. Otherwise, if these two virtual nodes are simulated by different physical nodes, then these two physical nodes must be directed connected or are two hops away from each other. Since each step of each phase is long enough for running \cseek twice, the two physical nodes simulating these two virtual nodes can still exchange one piece of information.

Lastly, notice that the maximum degree of $\mathcal{G}$ is $\Delta$, this implies for any pairs of neighboring nodes in $\mathcal{G}$, excluding the edge that connects them, there are at most $2\Delta-2$ other edges which are adjacent to them. Hence, in the line graph $\mathcal{G}_L$, the maximum degree is $2\Delta-2$, which implies $2\Delta$ colors is enough to color the nodes in $\mathcal{G}_L$.

We are now ready to prove the lemma. We first calculate the probability that a node $w_1$ becomes inactive after one phase.

Assume $w_1$ is still active at the beginning of phase $i$. If none of its neighbors are active at the beginning of phase $i$, then with probability $1/2$, node $w_1$ will choose a color that has not been used by any of its neighbors. (There is always some color available, as $w_1$ has at most $2\Delta-2$ neighbors while there are $2\Delta$ colors available initially.) Moreover, after this phase, $w_1$ will decide on this color and become inactive.

Otherwise, some neighbors of $w_1$ are still active at the beginning of phase $i$. Assume $W$ denotes this set of active neighbors. Consider a node $w_2\in W$. At the beginning of phase $i$, with probability $1/2$, node $w_1$ will choose a color that has not been used by any neighbors previously. Assume this event indeed happens. In such case, with probability $1/2$, node $w_2$ will choose not to pick any color in this phase. Otherwise, if node $w_2$ chooses a color, then with probability at most $1/k_{i,w_1}$, node $w_1$ and $w_2$ will choose identical colors. Here, $k_{i,w_1}$ denotes the number of colors that is available for $w_1$ to choose during phase $i$. As a result, we know conditioned on the event that $w_1$ chooses a color, the probability that $w_1$ chooses same color with some of its neighbors is at most $\sum_{w\in W}((1/2)\cdot(1/k_{i,w_1}))=\sum_{w\in W}{1/(2\cdot k_{i,w_1})}$. Notice that for $w_1$, the number of remaining available colors is always at least as large as the number of remaining active neighbors, thus we know $|W|\leq k_{i,w_1}$. Hence, we know conditioned on the event that $w_1$ chooses a color, the probability that $w_1$ chooses the same color with some of its neighbors is at most $1/2$.

At this time point, we can conclude that the probability that node $w_1$ becomes inactive after one phase is at least $(1/2)\cdot(1/2)=1/4$. As a result, we know in expectation, after each phase, at least $1/4$ fraction of remaining nodes will decide their colors and become inactive. Let $x_i$ be the number of active nodes at the beginning of phase $i$, and let $y_i$ be the number of nodes that turn from active to inactive after phase $i$. We know $\mathbb{E}(y_i)\geq x_i/4$. Now, according to the definition of expectation, we know $\mathbb{E}(y_i)\leq x_i\cdot\mathbb{P}(y_i\geq x_i/10)+(x_i/10)\cdot\mathbb{P}(y_i< x_i/10)\leq x_i\cdot\mathbb{P}(y_i\geq x_i/10)+x_i/10$. Hence, we know $x_i/4\leq x_i\cdot\mathbb{P}(y_i\geq x_i/10)+x_i/10$. This implies $\mathbb{P}(y_i\geq x_i/10)\geq 3/20$, which is saying after each phase, with some constant probability, a constant fraction of remaining nodes in $\mathcal{G}_L$ will decide their colors and become inactive. Since each phase is independent, by using a Chernoff bound, we know after $\Theta(\lg{n})$ phases, all nodes in $\mathcal{G}_L$ will decide and become inactive, w.h.p.

Finally, notice that after the $\Theta(\lg{n})$ phases of coloring, we have one additional execution of \cseek so that for each virtual node, the physical node simulating this virtual node will tell the color of the virtual node to the other physical node, w.h.p. By now, we have proved the lemma.
\end{proof}

\begin{proof}[\bf{Proof of Theorem \ref{thm-cgcast}}]
To prove the theorem, we only need to prove the correctness of the message dissemination part of \cgcast.

For a node $u$, we claim that if at the beginning of phase $i$ at least one of its neighbors knows the message that needs to be disseminated, then by the end of phase $i$, node $u$ will know this message, w.h.p. To see this, assume at the beginning of phase $i$, node $v$, which is a neighbor of $u$, already knows the message. Then, according to Lemma \ref{lemma-gb-node-color} and the protocol, during phase $i$, there must exist a step during which both $u$ and $v$ will go to the same channel (i.e., the dedicated communication channel). For each round within that step, $u$ will listen on the specified channel. Moreover, at least one node (which is $v$), and at most $\Delta$ nodes will broadcast the message on the specified channel. Since within each round there is a back-off procedure containing $\lg{\Delta}$ slots, we know after each round $u$ has at least a constant probability to obtain the message. Since each step contains $\Theta(\lg{n})$ rounds, we know after this particular step, $u$ will learn the message, w.h.p.

The above analysis shows that the message will propagate at least one hop per phase, w.h.p. Since the network diameter is $D=O(n)$, we know after $D$ phases, all nodes will obtain the message, w.h.p.
\end{proof}

\begin{proof}[\bf{Proof of Lemma \ref{lemma-nd-lower-bound-reduction}}]
We construct our player $\mathcal{P}_{\mathcal{A}}$ to simulate a network containing two nodes: $u$ and $v$. Assume $u$ and $v$ share exactly $k$ channels. Moreover, let $A=\{a_1,a_2,\cdots,a_c\}$ be $u$'s channel set and $B=\{b_1,b_2,\cdots,b_c\}$ be $v$'s channel set. Since we consider local channel label model, assume $u$ sees each $a_i$ labeled as $i$, and $v$ sees each $b_i$ labeled as $i$.

The first key observation for our simulation is that a $k$-matching $M$ over the complete bipartite graph with bipartition $(A,B)$ also describes a valid overlap of $k$ channels between $u$ and $v$. The second key observation for our simulation is that in order for $\mathcal{A}$ to solve neighbor discovery in such a network, there must exist a time slot in which $u$ lands on a shared channel with $v$ (until this happens, the neighbor discovery procedure makes no progress). When the overlaps are viewed as a matching, this is the same as saying that there must exist a time slot in which $u$ chooses some $a_i$ and $v$ chooses some $b_j$ such that $(a_i,b_j)\in M$.

We are now ready to describe our simulation and establish its correctness. The player $\mathcal{P}_{\mathcal{A}}$ simulates $u$ running with channel set $A$, and $v$ running with channel set $B$, and the (unknown to the player) matching $M$ chosen by the referee for this execution defining the $k$ overlapping channels. In each simulated time slot $r$, player $\mathcal{P}_{\mathcal{A}}$ guesses $(a_r,b_r)$, where $a_r$ is the channel selected by $u$ in this time slot, and $b_r$ is the channel selected $v$ in this time slot. If a guess does not win the game, it follows that $u$ has failed to land on a shared channel with $v$, so $\mathcal{P}_{\mathcal{A}}$ can correctly complete the simulated time slot by simulating no communication between $u$ and $v$.

As noted, to complete neighbor discovery, there must be a time slot in which $u$ shares a channel with $v$. In that simulated time slot, $\mathcal{P}_{\mathcal{A}}$'s guess will win the bipartite hitting game. Because $\mathcal{P}_{\mathcal{A}}$ gets at most one guess per simulated time slot, we can immediately have our lemma.
\end{proof}

\begin{proof}[\bf{Proof Sketch of Theorem \ref{thm-gb-lower-bound}}]
We first consider the $\Omega(c^2/k)$ part. The argument for it is similar to the one we used when proving lower bound for neighbor discovery. In particular, we can reduce the $(c,k)$-bipartite hitting game discussed earlier to two node broadcast (i.e., one source node and only one uninformed node) by demonstrating a strategy for using a fast (two nodes) global broadcast algorithm to create a fast solution to the $(c,k)$-bipartite hitting game. The details for this proof is very similar to the proof for Lemma \ref{lemma-nd-lower-bound-reduction}.

We then consider the $\Omega(D\cdot\min\{c,\Delta\})$ part. To see this, imagine the network graph $\mathcal{G}$ is a complete tree. Except the leaves, each other node in the tree has $\min\{c,\Delta\}-1$ children, and the hight of the tree is $D$. Notice, since each node has access to $c$ channels, and the maximum degree is $\Delta$, such tree can always be constructed. Assume the root node has a message that needs to be disseminated to all other nodes. Also assume any two siblings do not share any communication channels. Now, consider the $\min\{c,\Delta\}-1$ children of the root node $u_0$. Clearly, to send the message to all of them will take at least $\min\{c,\Delta\}-1=\Theta(\min\{c,\Delta\})$ time slots, as in each time slot node $u_0$ can only send the message to at most one child. W.l.o.g, assume $u_1$ is the last child of $u_0$ that receives the message from $u_0$, and it receives the message at time slot $\min\{c,\Delta\}-1$. Now, consider the $\min\{c,\Delta\}-1$ children of $u_1$. Again, we can show that it takes at least $\min\{c,\Delta\}-1=\Theta(\min\{c,\Delta\})$ times slots for all children of $u_1$ to get the message, and the last child of $u_1$ that receives the message can receive it no earlier than time slot $2(\min\{c,\Delta\}-1)$ since the beginning of execution. We can continue this process, and prove by induction that for nodes that are $d$ hops away from the root, it takes at least $d\cdot(\min\{c,\Delta\}-1)$ time slots for all of them to receive the message. Since the hight of the tree is $D$, we can obtain the $\Omega(D\cdot\min\{c,\Delta\})$ bound as desired.
\end{proof}

\end{document}